\documentclass[         
aps,                    
prd,                    
showpacs,               
nofootinbib,            
twocolumn,             %
showkeys,               %
preprintnumbers,        %
floatfix]               
{revtex4}               
\usepackage{graphicx,longtable}
\usepackage{color}
\usepackage[normalem]{ulem}
\usepackage{hyperref}
\begin{document}

\title{Turbulence effects on supernova neutrinos}

\author{James Kneller}
\email{jim_kneller@ncsu.edu}
\affiliation{Institut de Physique Nucl\'eaire, F-91406 Orsay cedex, CNRS/IN2P3 and University of Paris-XI, France} 
\affiliation{Department of Physics, North Carolina State University, Raleigh, NC 27695, USA}
\author{Cristina Volpe}
\email{volpe@ipno.in2p3.fr}
\affiliation{Institut de Physique Nucl\'eaire, F-91406 Orsay cedex, CNRS/IN2P3 and University of Paris-XI, France}
\today

\begin{abstract}
Multi-dimensional core-collapse supernova simulations exhibit turbulence of large amplitude and over large scales. As neutrinos pass through the supernova mantle 
the turbulence is expected to modify their evolution compared to the case where  
the explosion is free of turbulence.  
In this paper we study this turbulence effect upon the neutrinos modelling the turbulence expected from 
multi-dimensional simulations by adding matter density fluctuations to density profiles taken from one-dimensional hydrodynamical simulations. 
We investigate the impact upon the supernova neutrino transition probabilities as a function
of the neutrino mixing angle $\theta_{13}$ and turbulence amplitude.
In the high (H) resonant channel and with large $\theta_{13}$ values we find that turbulence
is effectively two flavor for fluctuation amplitudes $\lesssim 1\%$ and have 
identified a new effect due to the combination of turbulence and multiple H resonances 
that leads to a sensitivity to fluctuations amplitudes as small as $\sim 0.001\%$.  
At small values of $\theta_{13}$, beyond the range achievable in Earth based
experiments, we find that turbulence leads to new flavor transient effects in
the channel where the MSW H resonance occurs. 
Finally, we investigate large amplitude fluctuations which lead to three
flavor effects due to broken HL factorization and significant non-resonant
transitions and identify two non-resonant turbulence effects, one depending on
the $\theta_{13}$, and the other independent of this angle and due to the low (L) MSW resonance. 

\end{abstract}


\maketitle


\section{Introduction}

Core-collapse supernova simulations have reached a high degree of sophistication
and complexity in the last decade.
A new picture is emerging that shows how 
several ingredients might be necessary to achieve successful explosions. 
These include fluid instabilities (the SASI mode), realistic nuclear networks,
detailed neutrino transport, rotation, magnetic fields etc. 
\cite{Blondin:2003ApJ...584..971B,Mezzacappa:2005ju}. 
The flavor evolution of the neutrino signal from the next Galactic supernova as
a function of time and energy has the potential 
to reveal invaluable information about the explosion of the star and is being
thoroughly investigated for the case of spherically 
symmetric explosions.

Fully understanding neutrino flavor conversion in 
the supernova environment and the complex interplay with the various features of
the supernova dynamics is clearly a theoretical challenge.
Nevertheless, there has been significant ongoing progress and
recent developments have highlighted a variety of novel flavor conversion
phenomena. This is due to the inclusion of both
the neutrino-neutrino interaction and the use of dynamic density profiles taken
from simulations. While the former induces collective
effects \cite{Duan:2006an,Hannestad:2006nj,Balantekin:2006tg,Raffelt:2007xt,Gava:2008rp,
Dasgupta:2009mg} the latter produces dynamic multiple Mikheev-Smirnov-Wolfenstein (MSW) resonances which can lead to phase effects
\cite{Schirato:2002tg,Takahashi:2002yj,Fogli:2003dw,Tomas:2004gr,Fogli:2004ff,
Kneller:2005hf,Dasgupta:2005wn,Choubey:2006aq}. For a review see \cite{Duan:2009cd,Duan:2010bg}.
The first calculation for the supernova neutrino signal implementing both of
these effects together consistently is performed in \cite{Gava:2009pj} while the
impact of collective and dynamic MSW effects upon the diffuse supernova neutrino
background can be found in \cite{Chakraborty:2008zp,Galais:2009wi}. 

Multi-dimensional supernova simulations indicate strong turbulence,
particularly in the region between the reverse and forward shocks, generated by non-radial fluid flow
through the distorted shock surfaces \cite{2006A&A...453..661K,2006A&A...457..963S,2009Nonli..22.2775G} and other
multi-dimensional phenomena. But due to the absence of suitable multi-dimensional simulations 
important properties of the turbulence, such as its amplitude at late times, are unknown.
Turbulence effects upon neutrinos, not just in supernova, have been investigated in several 
works \cite{1990PhRvD..42.3908S,1996NuPhB.472..495N,1996PhRvD..54.3941B,1998PhRvD..57.3140H}. 
In \cite{Loreti:1994ry} neutrino oscillations in noisy media with
$\delta$-correlated fluctuations is considered with applications to a fluctuating matter density and magnetic
fields in the sun.  
In \cite{Loreti:1994ry} it is also found that depolarization can also occur even
if the fluctuations never cause the density to cross the neutrino resonance density i.e. 
they are off-resonance. The application to supernovae is studied in  \cite{Loreti:1995ae} with a similar
model for the turbulence to determine the implications for neutrino flavor
conversion and supernova dynamics (r-process and shock reheating).
In \cite{Fogli:2006xy} the authors use  the same model of delta-correlated
fluctuations for core-collapse supernovae, but now with profiles containing both
front and reverse shocks. A sensitivity to the third neutrino mixing angle is shown in the channels where
no MSW H resonance occurs (electron neutrino channel in inverted hierarchy and
electron anti-neutrino channel in normal hierarchy). 
Finally arguments were made in \cite{Friedland:2006ta} about the application of 
Kolmogorov correlated fluctuations and linear profiles to supernova. 
Regardless of their differences, all these studies agree that the inclusion of matter density fluctuations in media has
the possible consequence of producing completely mixed states (a process called `depolarization') if the 
amplitude is sufficiently large. At the same time all these
investigations are based on the solutions of differential equations for the 
averaged density matrix using just two neutrino flavors. The first full three flavor study of turbulence effects is performed in
Ref.\cite{Kneller:2010ky} where turbulence is superimposed upon profiles from
one-dimensional hydrodynamical simulations using a Kolmogorov spectrum for the
density fluctuations. In contrast to the previous work, here turbulence effects are studied by building a statistical
ensemble of instantiations for the neutrino survival probability. It is also
shown there that HL factorization might be broken in the limit of strong
turbulence. 
But despite the acknowledgment that turbulence may be important the 
effect is not considered in the calculations for the neutrino signal. Thus it 
remains unclear if the presence of turbulence will remove the information about the
explosion imprinted in the neutrino signal and which is supposed to indicate the missing 
information about neutrino mixing. 

The purpose of this paper is to revisit the subject of turbulence effects upon 
core-collapse supernova and explore the effects of changing the angle $\theta_{13}$ and the turbulence
amplitude. As in Ref. \cite{Kneller:2010ky} we follow the neutrino evolution in 
the star using matter density profiles from a one-dimensional hydrodynamical
simulation with Kolmogorov correlated turbulence superimposed. Similarly we construct an
ensemble of results computed using a phase-retaining 3-flavor integrator instead
of determining the average density matrix and survival probabilities. We then
investigate the distributions of the probabilities and compute the mean and the
variance in order to study how the effects we find evolve as a function of the 
parameters. Our paper is structured as follows: in Section II the model and the numerical
procedures are described, Section III presents the numerical results of turbulence effects on the survival
probabilities first for the case of small amplitude turbulence and both large and small $\theta_{13}$ and then 
large amplitude fluctuations which can cause three flavor mixing in both the
H-resonant and non-H-resonant channel, Section IV contains our conclusions.

\section{The model}

Neutrinos evolve through matter according to the Schroedinger-like equation:
\begin{equation}
\imath\frac{d}{dt}|\psi\rangle = H\,|\psi\rangle = (K + V)\,|\psi\rangle.
\label{eq:1}
\end{equation}
In the absence of neutrino-neutrino interactions the Hamiltonian is composed of two terms: the vacuum term $K$, and
the potential term $V$ due to the matter. In the flavor basis the vacuum term $K^{(f)}$ depends
upon two mass-square differences $\delta m^2_{12}$ and $\delta m^2_{23}$ and on the
parameters of the Maki-Nakagawa-Sakata-Pontecorvo matrix 
which are the three neutrino mixing angles usually labeled as $\theta_{12}$,
$\theta_{13}$, $\theta_{23}$ and three phases, one Dirac and two Majorana.
For all results in this paper we set the oscillation frequencies and angles to
$\delta m^2_{12}= 8 \times 10^{-5}$eV$^2$, $|\delta m^2_{23}|= 3 \times
10^{-3}$eV$^2$ and by $\sin^{2} 2\theta_{12}=0.83$ and $\sin^{2} 2\theta_{23}=1$
\cite{Amsler:2008zzb}. We shall consider both normal and inverse hierarchies and
various values for the unknown angle $\theta_{13}$ up 
to its limit from Chooz \cite{1999PhLB..466..415A,2000PhLB..472..434A}. We do
not consider any effect coming from the Dirac CP phase $\delta$. For a
discussion of its impact see \cite{Balantekin:2007es,Gava:2008rp}.

The potential term, at some fixed time is diagonal in the flavor basis i.e.\
$V^{(f)}(\mathbf{r})=diag(V_{e}(\mathbf{r}),V_{\mu}(\mathbf{r}),V_{\tau}(\mathbf
{r}))$ with $V_{e}(\mathbf{r}) =\sqrt{2}\,G_{F}\,n_e(\mathbf{r})$ and
$n_e(\mathbf{r})$ the electron density. 
The potentials $V_{\mu}(\mathbf{r})$ and $V_{\tau}(\mathbf{r})$ are negligible
compared to $V_{e}(\mathbf{r})$ and will be ignored hereafter.
The turbulence we shall study in this paper enters into equation (\ref{eq:1}) through $n_e$ 
so in order to compute the neutrino evolution one needs to
provide a density profile from which we can construct $V_{e}$. 
Ideally a study of the effect of turbulence in supernova upon the neutrinos
would use density profiles taken from successfull multi-dimensional
core-collapse supernova simulations. While significant progress 
continues to be made in the sophistication of the simulations
we are still far from being able to extract
fluctuation characteristics (scale, power spectra and amplitude) from 
them at late times and in the outer regions of the supernova where our interest
lies. Therefore one must adopt a pragmatic approach by using a `turbulence' free explosion and then adding the
turbulence to it in some prescribed fashion \cite{Fogli:2006xy,Kneller:2010ky}.
The scale, power spectrum and amplitude of the turbulence are thus variable.
While this approach requires calibration before it can be applied
to understanding turbulence in multi-dimensional supernova simulations, an exploration of the turbulence
parameter space can discover general features of turbulence effects upon the
neutrinos. There are also several advantages of this approach: first, the
underlying profile is the same so that non-turbulent features - such as the
shocks - are always located at the same positions, second the turbulence in the
profiles of the members of an ensemble is uncorrelated, and third, if the
turbulence has a zero expectation value then the average potential $\langle
V\rangle$ is known exactly allowing us to compare the effects of turbulence to the 
situation when turbulence is absent.

Now that we have outlined our approach we spell out the details. We
start with a profile taken from a spherically symmetric simulation, i.e.\ a one
dimensional model. We shall ensure that this one dimensional profile is also the
average profile $\langle V\rangle(r)$ once the turbulence is added.
The one-dimensional density profile we employ is the $4.5\;{\rm s}$ snapshot taken
from the hydrodynamical simulation of Ref.\cite{Kneller:2007kg} with $Q=3.36
\times 10^{51}$ erg. This particular simulation was chosen because it has the greatest resemblance
with the 2D hydrodynamical simulations there and because the density of the region
between the shocks - into which we will add the turbulence - overlaps with 
the H resonance density for neutrino energies between $10$ and $80$ MeV as shown
in Galais \emph{et al.} \cite{Galais:2009wi}.
Although we show results for a given snapshot in time our findings are
valid for the entire period the shock wave is in the H resonance region for neutrino
energies of order $\mathcal{O}(10\;{\rm MeV})$ i.e.\ from $t
\sim 2\;{\rm s}$ to $t\sim 10\;{\rm s}$ or so.
Later the shocks will move to affect the L resonance but the
neutrino flux will be so small at these times that turbulence effects
may be well-nigh impossible to observe. The turbulence is included in the model by writing the potential 
term in Eq.(\ref{eq:1}) as
\begin{equation}\label{eq:2}
V(r) = (1+F(r))\,\langle V\rangle(r)
\end{equation}
with $F(r)$ a random field. The field $F(r)$ is constructed as
\begin{eqnarray}
F(r)&=&\frac{C_{\star}}{\sqrt{N_k}}\,\tanh\left(\frac{r-r_r}{\lambda}\right)\,
\tanh\left(\frac{r_s-r}{\lambda}\right)\nonumber \\
&& \times\sum_{n=1}^{N_k}\left\{ A_{n} \cos\left[k_{n}\,(r-r_r)\right] + B_{n}
\sin\left[k_{n}\,(r-r_r)\right] \right\}\nonumber \\
&& \label{eq:3}
\end{eqnarray}
for $r_r \leq r \leq r_s$ and is zero outside this range. In this equation the
parameter $C_{\star}$ sets the amplitude of the fluctuations. The two radii
$r_r$ and $r_s$ are the positions of the reverse and forward shock respectively;
the two $\tanh$ terms are included to suppress fluctuations close to the shocks and 
prevent discontinuities at $r_s$ and $r_r$, and the parameter $\lambda$ is a scale over which the fluctuations reach their
extent size. We set $\lambda=100\;{\rm km}$. The second half of equation
(\ref{eq:3}) is the discrete Fourier representation of a random field. The
members of the set of co-efficients $\{A\}$ and $\{B\}$ are independent standard
Gaussian random variates with zero mean thus ensuring the vanishing expectation
value of $F$. To determine the $N_k$ $k$'s, $A$'s and $B$'s for an instantiation
of $F$ the `Randomization method' described in Ref.\cite{MK1999} is employed.
The number of modes used throughout most of this paper is $N_k=100$ but to assure the reader that this is sufficient 
we shall also compute one of our principal results using $N_k=1000$ so that the reader can observe that the results for $N_k=100$ are essentially similar. 
Finally, the wavenumber cutoff $k_{\star}$ is set to $k_{\star}=\pi/(r_s-r_r)$ i.e.\ a wavelength twice the distance
between the shocks. 
The power spectrum $E(k)$ we use is
\begin{equation}\label{eq:4}
E(k) = (\alpha-1) \left( \frac{k_{\star}}{k} \right)^{\alpha}.
\end{equation}
Throughout this work we shall adopt a Kolmogorov spectrum where $\alpha=5/3$. 
Note that the power spectrum in Eq.(\ref{eq:4}) is very different from the one
of the $\delta-$correlated power spectra of Refs.\cite{Loreti:1995ae} and \cite{Fogli:2006xy}.

Now that all terms in Eq.(\ref{eq:1}) are defined, we solve the three flavor
neutrino evolution equation with the phase-retaining integrator of
Ref.\cite{2009PhRvD..80e3002K}.  
The quantities we are interested in calculating, are the neutrino and
antineutrino survival and appearance probabilities, i.e.\ $P(\nu_j\rightarrow\nu_i)$ and
$P(\bar{\nu}_j\rightarrow\bar{\nu}_i)$, in the matter basis since the 
flux of neutrinos emerging from the star is given by these quantities,
multiplied by the appropriate initial fluxes at the proto-neutron star. The
relationship between the flavor and matter bases can be found in Kneller \& McLaughlin
\cite{2009PhRvD..80e3002K}. Such probabilities can be formed from the wavefunction for each initial pure
state or from the elements of the $S$-matrix.
For brevity we shall often use the notation $P_{ij}=P(\nu_j\rightarrow\nu_i)$
and $\bar{P}_{ij}=P(\bar{\nu}_j\rightarrow\bar{\nu}_i)$.

\section{Numerical results: Turbulence impact on the probabilities}

Our aim is to explore the parameter space of our model in order to survey the
various effects turbulence produces and 
derive the relationships between the survival and crossing probabilities as a
function of the parameters. This represents a large undertaking, for now 
we shall limit our parameter space to the mixing angle $\theta_{13}$ and the
fluctuation amplitude $C_{\star}$ and shall reserve for a later date \cite{JKCVnext} an exploration of the effects of changes in the 
turbulence power spectrum, the snapshot profile and the turbulence location. In particular we shall not consider the effect of turbulence in front of the forward shock, 
an issue intimately tied to the observability of neutrino transformation effects 
in the neutrino burst from the next galactic supernova.
   
We begin with the case of large $\theta_{13}$ and explore the effects in the H resonant
channel before considering smaller values of $\theta_{13}$. We shall finish
with the effects of large amplitude fluctuations that can break HL factorization
and lead to large effects in the non-resonant channels.

\subsection{Effective two flavor mixing in the H resonant channel}
\begin{figure}[t]
\includegraphics[clip, width=\linewidth]{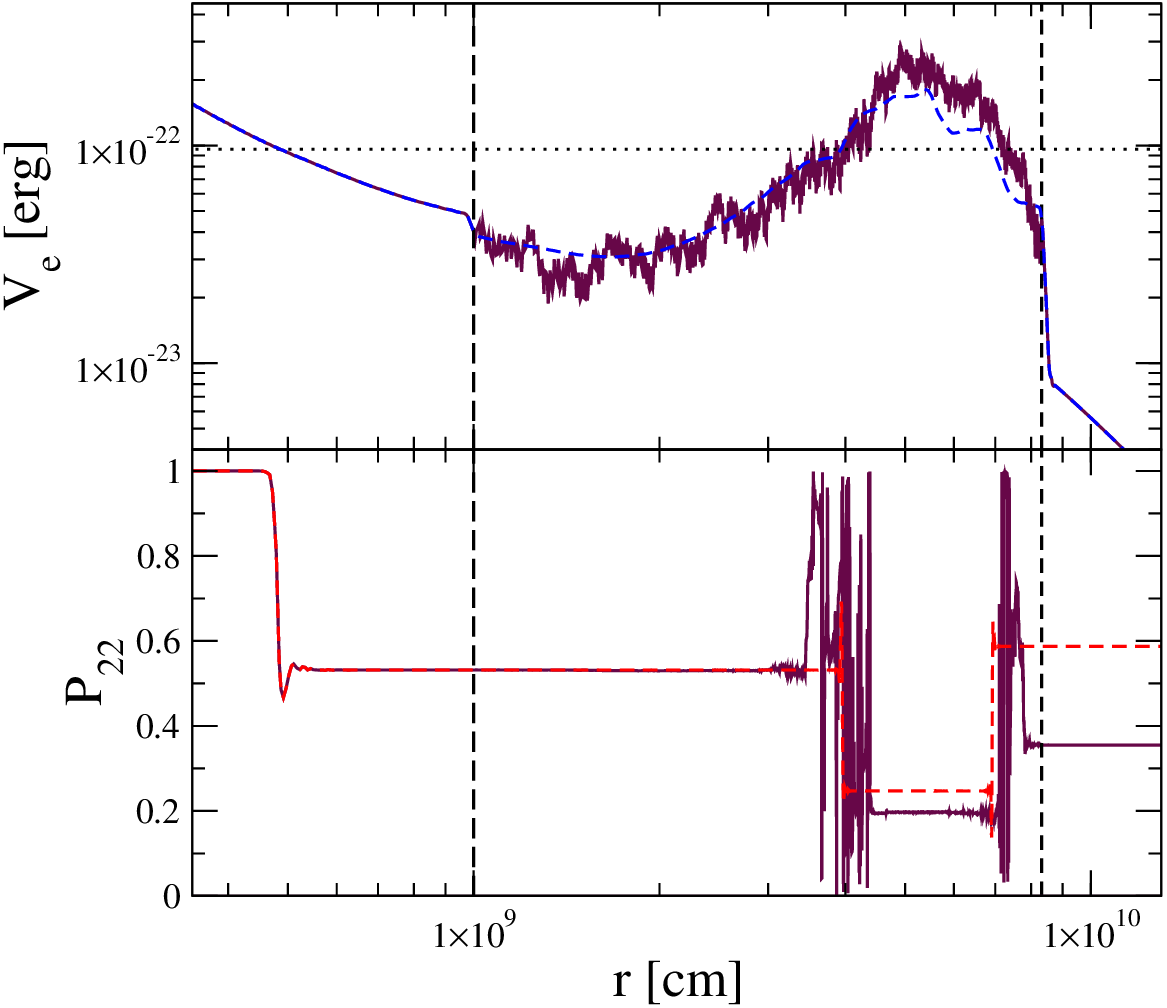}
\caption{One instantiation of the density profile (top panel) and the survival
probability $P_{22}=P(\nu_2\rightarrow\nu_2)$ for a $E=25\;{\rm MeV}$
neutrino (bottom panel) as functions of the radius $r$. The turbulence region is
identified by the vertical dashed lines in both panels, the H resonance density is the
horizontal line in the upper panel determined using the two flavor formula. The
underlying base profile is shown as the dashed line in the upper panel, the
evolution through the base profile as the dashed line in the lower panel. The
value of  $\theta_{13}$ is $\sin^{2}(2\theta_{13})=4\times 10^{-4}$, the
hierarchy is normal, and the amplitude of the density fluctuations is
$C_{\star}^2 =0.1$.\label{fig:1}}
\vspace*{2mm}
\end{figure}
\begin{figure}[t]
\vspace*{8mm}
\includegraphics[clip, width=\linewidth]{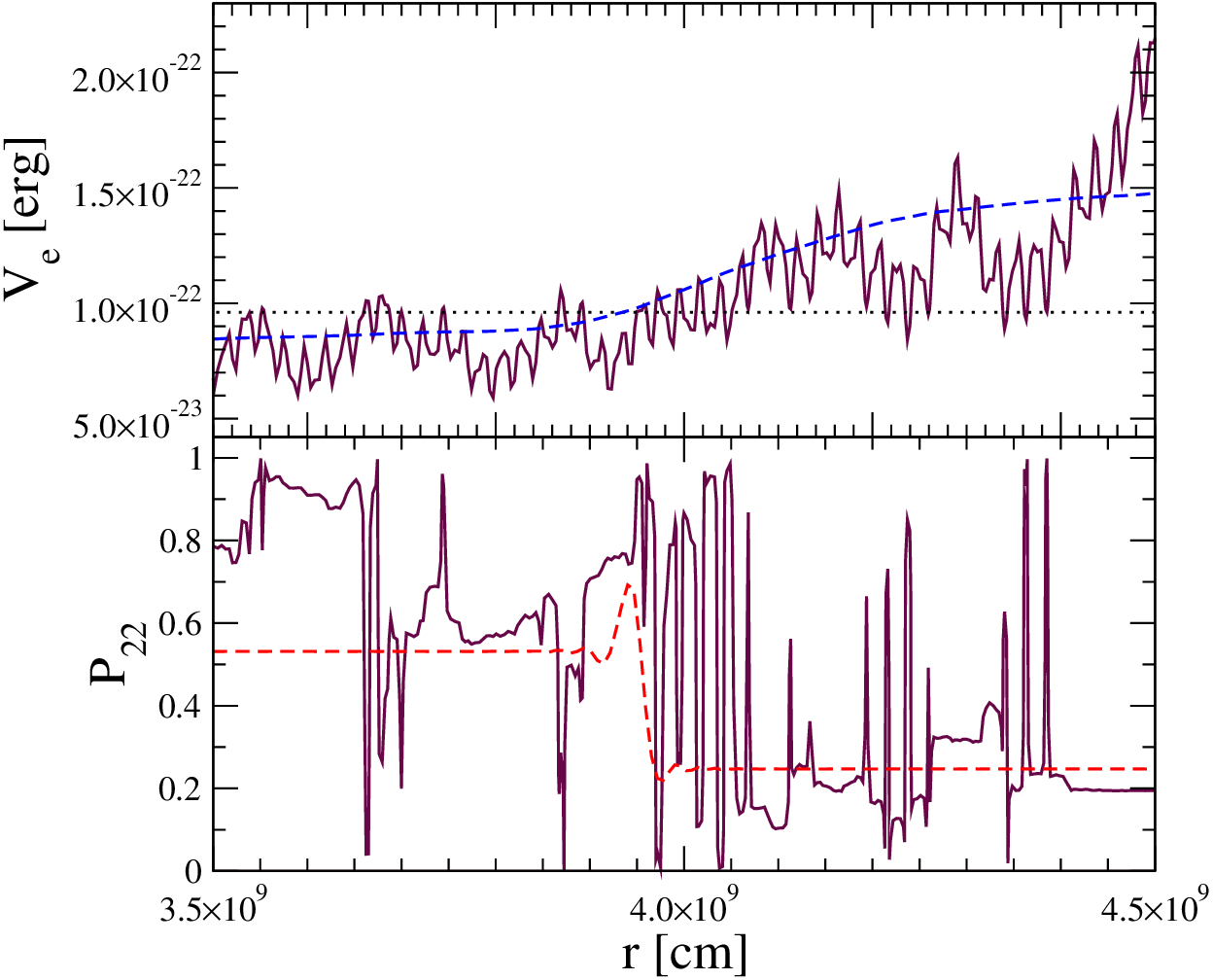}
\vspace*{5mm}
\caption{An enlargement of figure (\ref{fig:1}) to show the detailed neutrino
evolution in the MSW H resonance region.\label{fig:2}}
\vspace*{2mm}
\end{figure}
\begin{figure}[t]
\vspace*{2mm}
\includegraphics[clip, width=\linewidth]{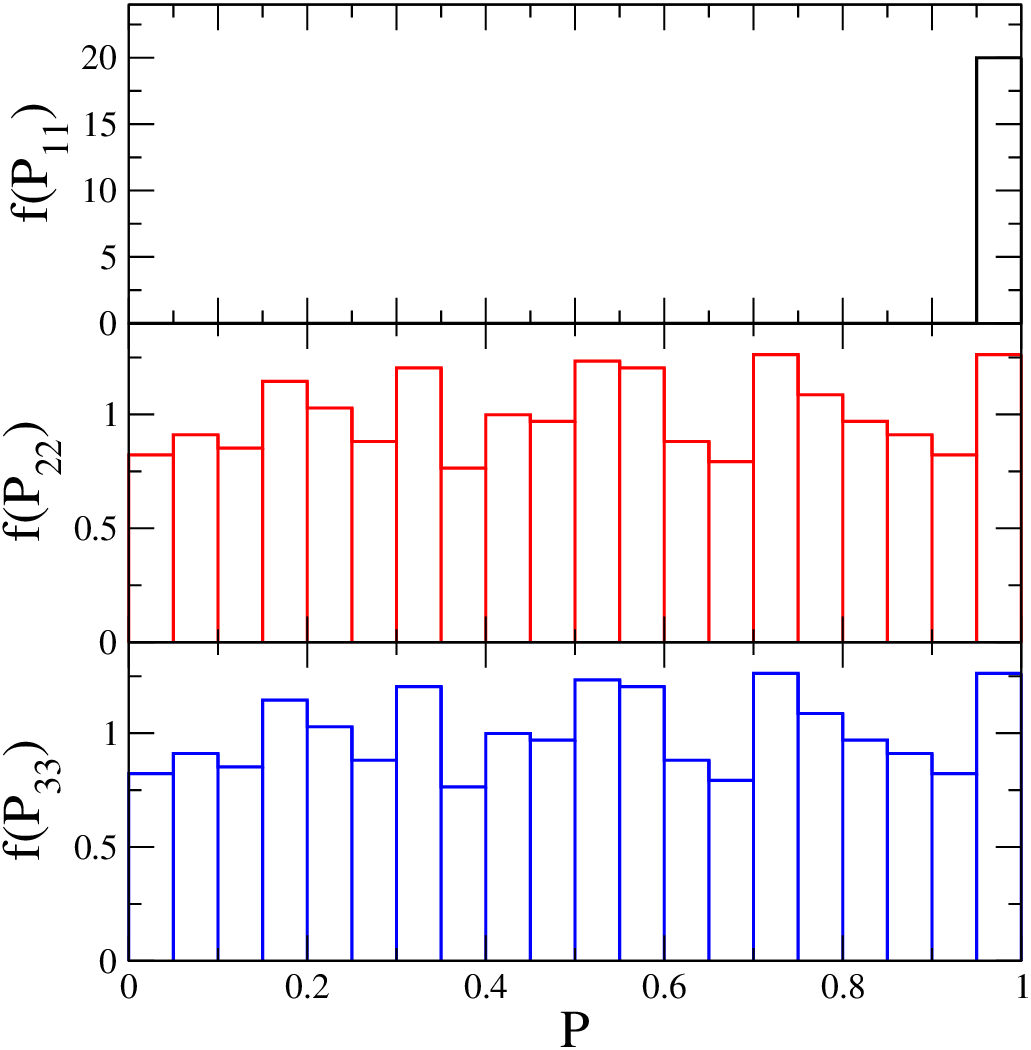}
\vspace*{5mm}
\caption{A normalized frequency histogram of 1012 calculations of $P_{11}$ (top
panel)
$P_{22}$ (middle panel) and $P_{33}$ (bottom panel) for $E=25\;{\rm MeV}$
neutrinos. The hierarchy is normal, $\sin^{2}(2\theta_{13})=4\times 10^{-4}$ and
 $C_{\star}^2 = 0.01.$ \label{fig:3}}
\vspace*{2mm}
\end{figure}
By following the neutrino as it progresses through the profile for a given
instantiation we discover that the effect of turbulence can span a range from
the perturbative to the strong. First, turbulence introduces
many new H resonances which lead to 
substantial changes from the evolution of the same neutrino through the profile
sans turbulence. We also observe the perturbative effect caused by
non-resonant fluctuations i.e.\ those fluctuations of the density that approach,
but do not intersect, the resonance density.
These two limits, the strong and the perturbative, can be seen in
figure (\ref{fig:1}) where we show the evolution 
of the second matter eigenstate survival probability,
$P_{22}$ (lower panel), as a function of the radius $r$ through one
instantiation of the matter density profile (top panel). The results are for a
`large' value \cite{Dighe:1999bi} of $\theta_{13}$ corresponding to
$\sin^{2}2\theta_{13}=4\times 10^{-4}$, a $25\;{\rm MeV}$ neutrino and the normal
hierarchy. We see that without fluctuations the neutrino experiences just three
H resonances for this chosen profile and energy, and one is located outside the
turbulence region. 
After the inclusion of turbulence the number of MSW resonances increases
significantly. Since such resonances are semi-adiabatic for this value of
$\theta_{13}$ the presence of the extra resonances due to turbulence causes the
neutrino survival probability to undergo large transitions as the neutrino
passes through them.
This can be seen more clearly in figure (\ref{fig:2}), an enlargement of figure (\ref{fig:1}). At each new H resonance caused by a downward fluctuation of the density 
the matter mixing angle
$\tilde{\theta}_{13}$ will swing from a value close to it's high density limit
of $90^{\circ}$ through $45^{\circ}$ towards its vacuum value and then back again: for an upward
fluctuation that crosses the resonance density the matter mixing angle does the
opposite swinging from close to the vacuum value up towards $90^{\circ}$ and then down
again. Either way, the non-adiabaticity parmeter $\Gamma_{23}$
\cite{2009PhRvD..80e3002K} - which describes the mixing between matter states
$\nu_2$ and $\nu_3$ - is proportional to the derivative
$d\tilde{\theta}_{13}/dr$\footnote{More exactly one observes that both the
$\Gamma_{23}$ and the $\Gamma_{13}$ non-adiabaticity parameters defined in
\cite{2009PhRvD..80e3002K} are proportional to  $d\tilde{\theta}_{13}/dr$ but as
a consequence of the evolution of the matter mixing angle $\tilde{\theta}_{12}$
in a normal hierarchy the non-adiabaticity function $\Gamma_{13}$ is suppressed.}.
$\Gamma_{23}$ is largest in the region of the resonance and if ever $\Gamma_{23}
\gtrsim 1$ then we have significant mixing between states $\nu_2$ and $\nu_3$
and thus a change in the survival probability $P_{22}$.
Similar results are obtained for the antineutrino matter eigenstate survival
probabilities $\bar{P}_{11}=P(\bar{\nu}_1\rightarrow \bar{\nu}_1)$ and
$\bar{P}_{33}=P(\bar{\nu}_3\rightarrow \bar{\nu}_3)$ in the case of
inverted hierarchy because now it is the antineutrino non-adiabaticity
$\bar{\Gamma}_{13}$ which is proportional to the derivative
$d\tilde{\theta}_{13}/dr$ and the H resonance 
mixes states $\bar{\nu}_1$ and $\bar{\nu}_3$.

Figure (\ref{fig:2}) also allows us to observe the perturbative effect from
non-resonant fluctuations. At non-resonant fluctuations the matter mixing angle
$\tilde{\theta}_{13}$ now varies over a smaller extent. The closer the
fluctuation approaches the resonance the larger the variation. Despite the
smaller variation of $\tilde{\theta}_{13}$ at a non-resonant fluctuation
compared to a resonant fluctuation 
the evolution of the neutrino state is
governed by the \emph{derivative} of this angle which can be large. For this
reason we understand why non-resonant fluctuations can lead to the evolution of
$P_{22}$ as observed.

The evolution of a neutrino with a particular energy and some set of oscillation
parameters through the turbulent density profile is unique for each
instantiation of the turbulence. If we change the $F$ then we find a completely
different final state. If we repeat this exercise many times 
then we can construct an ensemble of final states from which we can study the
general net effect of the turbulence  i.e.\ the state of the neutrino after it
has passed through the entire turbulence region and exits the supernova. The
results of such an exercise are shown in figure (\ref{fig:3}) which displays the
survival probability distribution of the three neutrino matter eigenstates in
the case of $E=25\;{\rm MeV}$ neutrinos, a normal hierarchy, 
$\sin^{2}(2\theta_{13})=4\times 10^{-4}$ and $C_{\star}^2 = 0.01.$
Since mixing at the H resonance is between states $\nu_2$ and $\nu_3$ and also
the amplitude of the fluctuation is relatively small the survival probability
$P_{11}$ is a delta-function at $P_{11}=1$. Larger fluctuation amplitudes lead
to broken HL factorization \cite{Kneller:2010ky}. 
In contrast one can see that the distributions for $P_{22}$ and $P_{33}$ are
consistent with uniform and the average of the ensemble for these two
probabilities is $1/2$. This implies that full `depolarization' occurs for matter states $\nu_2$ and
$\nu_3$: one starts with a pure neutrino state and ends up with a flavor state
that is completely random so that all information of the initial state is lost
in the averages. 

\begin{figure}[t]
\includegraphics[clip, width=\linewidth]{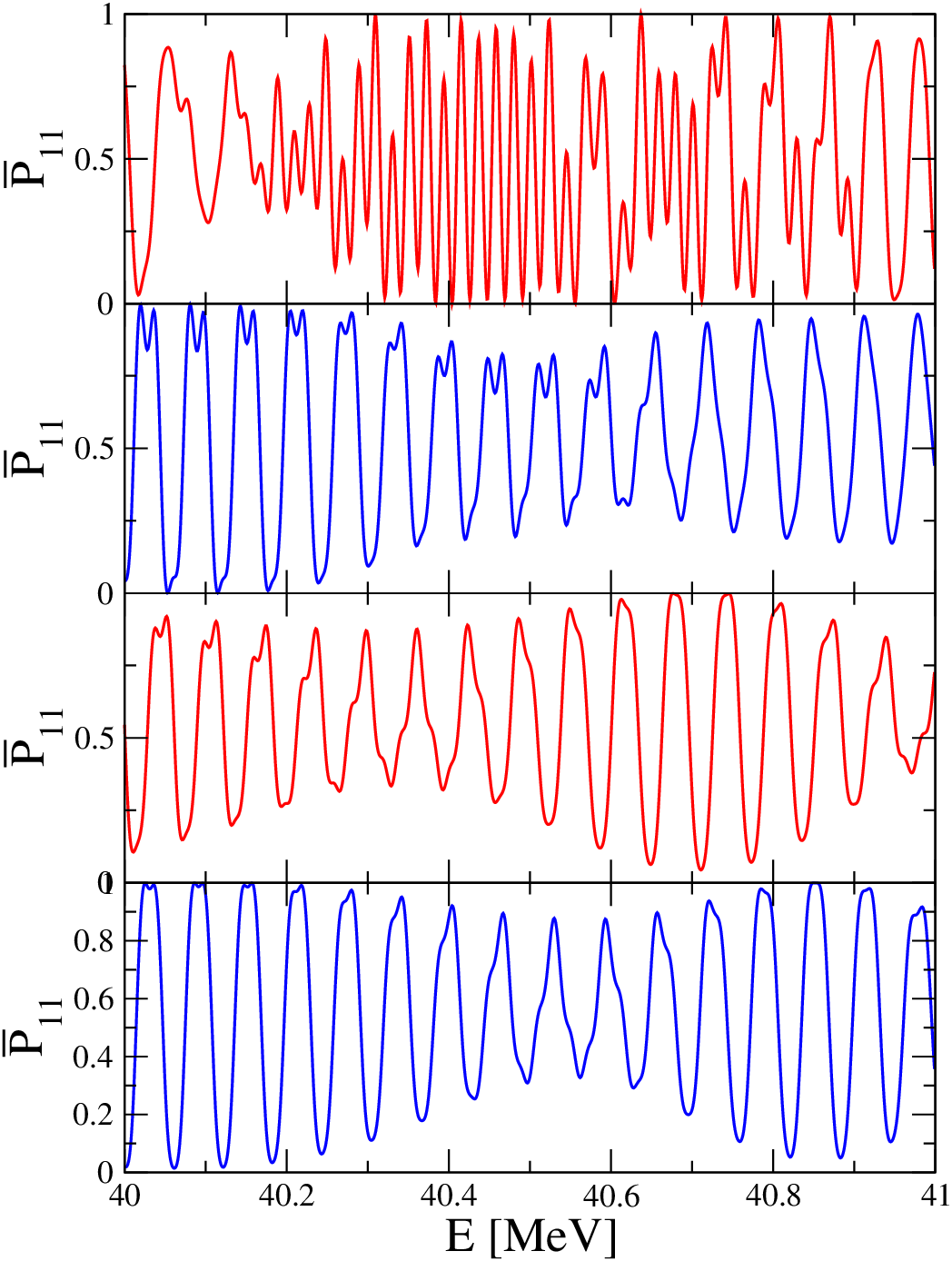}
\vspace*{5mm}
\caption{A very fine energy resolution of $\bar{P}_{11}$ between 40 and
41 MeV. The bottom panel is for no fluctuations, the next is
$C_{\star}=10^{-3}$, then $C_{\star}=10^{-2}$ and $C_{\star}=10^{-1}$. The
results correspond to inverted hierarchy and a `large' third neutrino mixing
angle given by $\sin^{2}(2\theta_{13})=4\times 10^{-4}$.
 \label{fig:4}}
\vspace*{3mm}
\end{figure}
As we decrease the fluctuation amplitude the number of new resonances created
tends to zero and also the perturbative effect from non-resonant fluctuations
disappears. But surprisingly we find that the turbulence does not disappear
quickly with $C_\star$. 
Turbulence continues to have a large impact upon the final state probabilities
one finds in an ensemble even for fluctuation amplitudes as small as
$C_{\star}\sim 10^{-5}$. We have investigated the reason for this extreme
sensitivity and have found that it is due to a change in the relative phase
difference between the extant resonances of the profile. 
That the sensitivity to small $C_{\star}$ is due to the combination of phase
effects and turbulence can
be deduced from figure (\ref{fig:4}). Here the hierarchy is inverted but this
makes no difference to the argument. Going from the lower to upper panels one
can see that for values of $C_{\star}=10^{-3}-10^{-2}$ the turbulence does not
change the correlation energy scale of $\sim 50\;{\rm keV}$. Only for
$C_{\star}\gtrsim 10^{-1}$ does one observe a decrease in the correlation energy
scale to $\sim 10\;{\rm keV}$ as the turbulence effects start dominating over
the phase effects. From the similarity of the middle panels of the figure with
the case of no fluctuations one concludes that 
the number of resonances must be the same for all of them. Once this is
appreciated one can then understand the  sensitivity to $C_{\star}$ from the
case of two resonance phase effects discussed in Kneller \& McLaughlin
\cite{Kneller:2005hf} and Dasgupta \& Dighe \cite{Dasgupta:2005wn}. 
Using a two-flavor approximation of the mixing between states $\alpha$ and
$\beta$ the crossing probability for a neutrino passing through two resonances
is 
\begin{eqnarray}
P_{C}&=&P_1\,(1-P_2)+(1-P_1)\,P_2 \nonumber \\
&&+ 2\sqrt{P_1\,P_2\,(1-P_1)(1-P_2)}\,\cos\Phi
\end{eqnarray} 
where $P_i(r_i)$ are the crossing probabilities for the resonances separately
and 
\begin{equation}
\Phi=\int_{r_1}^{r_2}\,dr (k_{\alpha}-k_{\beta})
\end{equation} 
is an interference term in which the $r$'s are the locations of the resonances
and the $k$'s are the matter eigenvalues. 
Turbulence leads to fluctuations of $k_{\alpha}-k_{\beta}$, relative to the case
without turbulence, and it also shifts the resonance positions $r_1$ and $r_2$.
But if the fluctuation amplitude is small the changes in the resonance positions
and the gradient there are also small and so $P_1$ and $P_2$ are essentially
unaltered. This leaves just the fluctuations of $Phi$ due to the small shifts in the 
resonance positions and the difference of the eigenvalues $k_{\alpha}-k_{\beta}$ between them.  
In order to generate a large effect we need a change $\delta \Phi \sim 1$ but this is typically 
small compared to $\Phi$ itself which is often $\Phi \gg 1$ because the resonances are very far apart.
If $\delta \Phi \sim 1$ then $P_C$ will vary over the range $\Delta P_C = 4\sqrt{P_1\,P_2\,(1-P_1)(1-P_2)}$ and one may show that
$\Delta P_C \leq 1/2$ i.e. less than the spread of the uniform distribution shown above. 

\begin{figure}[t]
\includegraphics[clip, width=\linewidth]{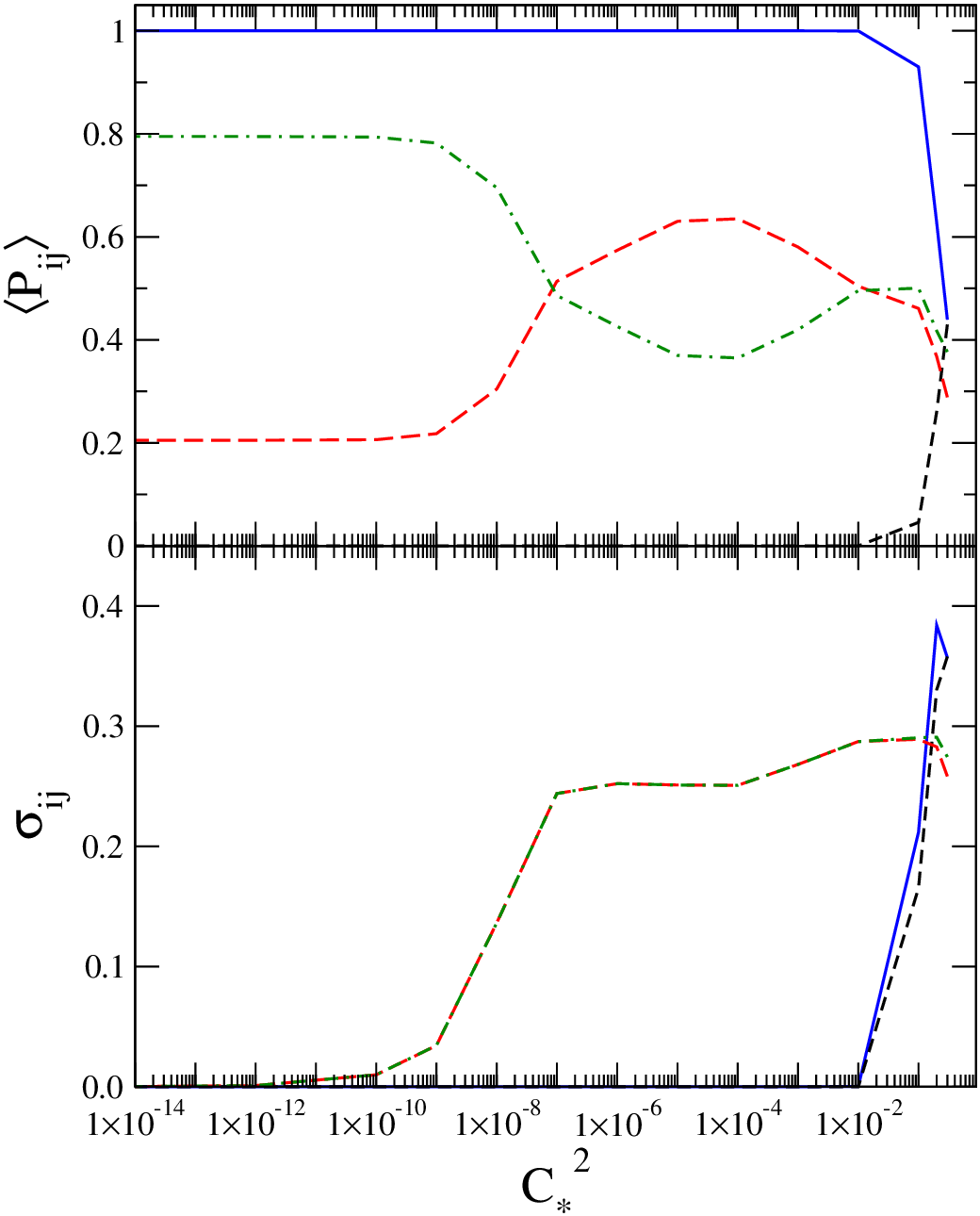}  
\caption{The mean and the variance of the neutrino matter survival probabilities
for a $E=60\;{\rm MeV}$ neutrino, as a function of the noise amplitude.
At every value of $C_\star$ these quantities are calculated from at least 1000
samples of the turbulence $F$. The mean and rms variance of $P_{11}$ is represented by the solid line in both
panels, $P_{22}$ is shown by the long-dashed, $P_{12}$ and $P_{23}$ are the
short-dashed and dash-dot lines respectively. The
hierarchy is normal and $\sin^{2}(2\theta_{13})=4\times 10^{-4}$.
The number of k modes is $N_k=100$. \label{fig:5}
}
\vspace*{2mm}
\end{figure}
\begin{figure}[t]
\includegraphics[clip, width=\linewidth]{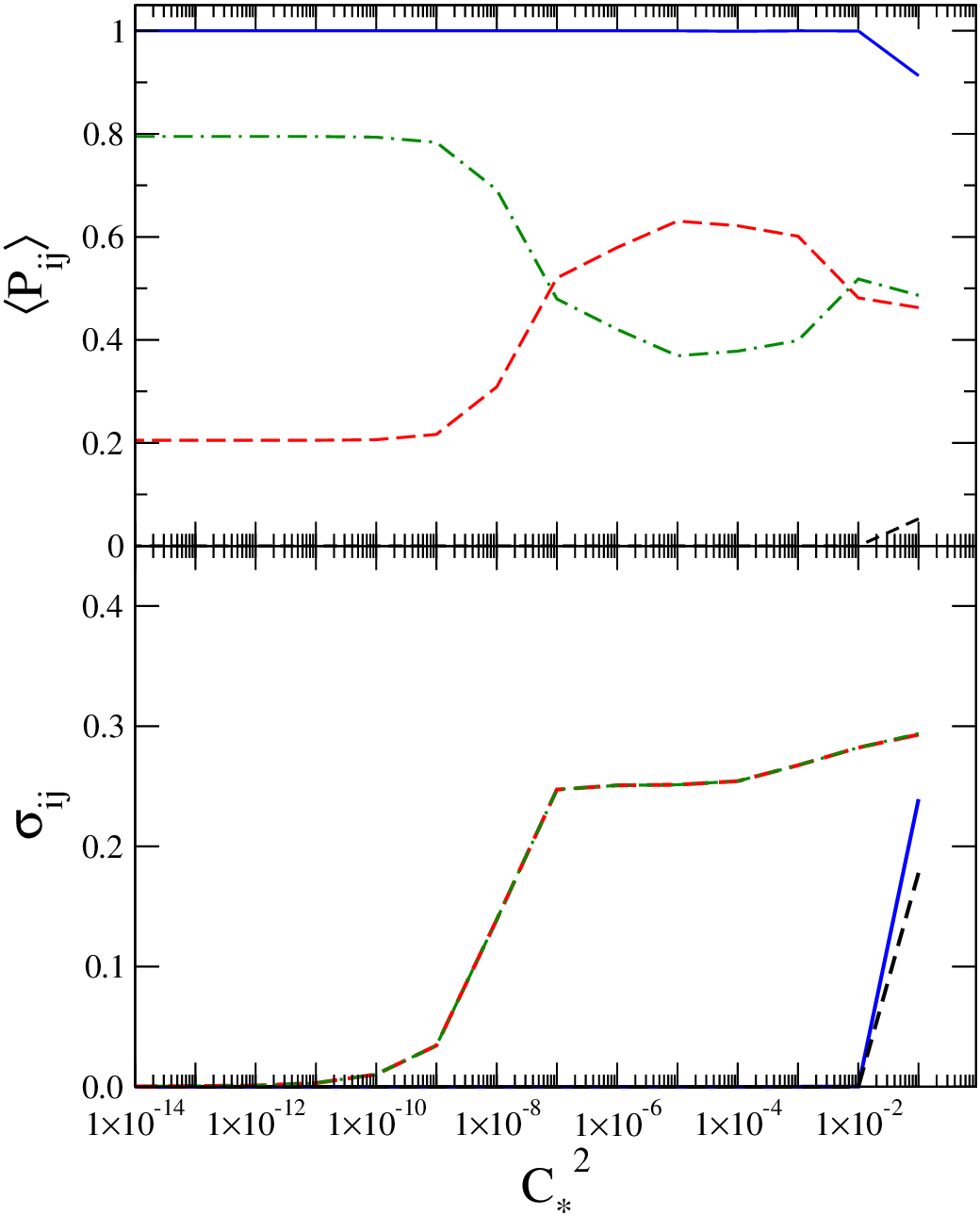} 
\caption{The same as figure (\ref{fig:5}) but with $N_k=1000$. \label{fig:6} }
\vspace*{2mm}
\end{figure}

So with this understanding of how neutrinos can be sensitive to such small amplitude fluctuations we present 
the full evolution of the survival and crossing
probabilities as a function of $C_{\star}$ for this case of a normal hierarchy
and `large' $\theta_{13}$. Figures (\ref{fig:5}) and (\ref{fig:6}) show plots of the probabilities $P_{11}$,
$P_{12}$, $P_{22}$ and $P_{23}$ as a function of the matter density fluctuation
amplitude for $60\;{\rm MeV}$ neutrinos in a normal hierarchy with
$\sin^{2}2\theta_{13}=4\times 10^{-4}$. The energy has changed compared to previous figures simply for the sake of clarity of the 
figure, the results are otherwise typical. And to demonstrate that the results are insensitive to 
the number of k modes we present results for $N_k=100$ in figure (\ref{fig:5}) and in figure (\ref{fig:6}) 
$N_k=1000$. From these figures four different ranges of $C_{\star}$ can be identified. For
values smaller than $C_{\star}^2 \lesssim 10^{-10}$ turbulence effects are
negligible and the probabilities one derives are always the same and determined
by the MSW resonances of the underlying adopted profile.
For values between $C_{\star}^2 \sim 10^{-10}$ and $10^{-2}$ the turbulence
effects and the phase effects induced by the multiple MSW resonances of the
underlying profile, that must occur even in the absence of fluctuations, combine
to produce broad variations over the ensemble but smaller than the
depolarization limit. In the figures one observes that the rms variance of $P_{22}$ and 
$P_{23}$ plateau at $\sigma =0.23$, whereas the two flavor depolarized limit would give
$\sigma=0.28$. In the region $10^{-2}$-$10^{-1}$ turbulence effects start
dominating over phase effects and lead to depolarization of states $\nu_2$ and
$\nu_3$ but not $\nu_1$. Note that figure (\ref{fig:3}) shows the distribution for
the survival probabilities when $C_{\star}$ is in this range and one observes in
figures (\ref{fig:5}) and (\ref{fig:6}) that $\sigma$ reaches $0.28$ for these values of $C_{\star}$. Finally,
for $C_{\star}^2 \gtrsim 0.1$ we start to transit to the case of three flavor
depolarization where all three matter states start to mix due to broken HL
factorization \cite{Kneller:2010ky}. The distributions start to become
triangular and the mean value drops towards $1/3$. This transition to three flavor mixing is 
studied more carefully in the next section.

\begin{figure}[t]
\vspace*{2mm}
\includegraphics[clip, width=\linewidth]{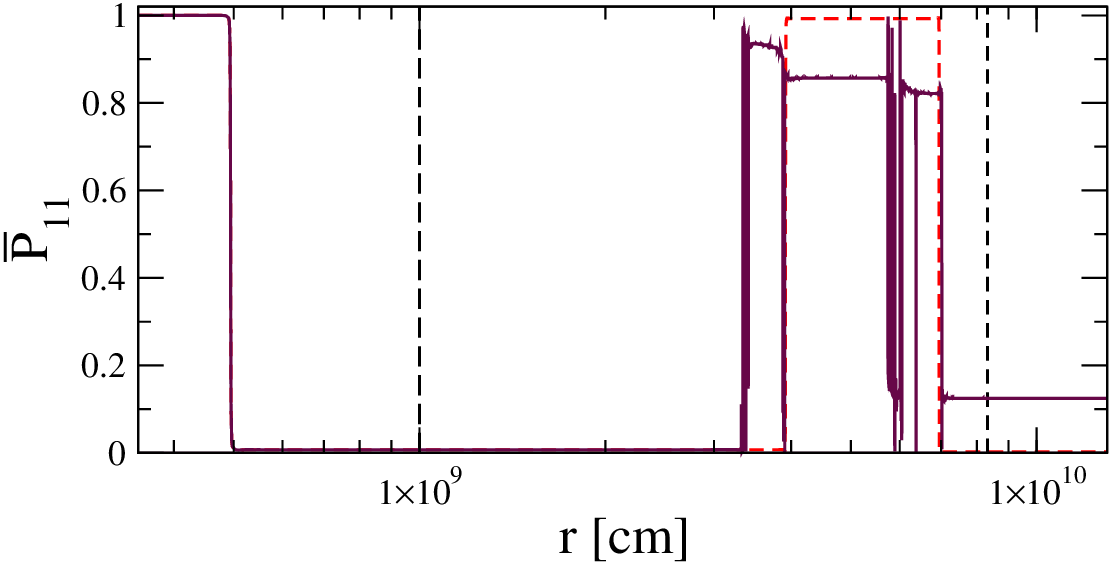}
\caption{The survival probability $\bar{P}_{11}$ for a
$E=25\;{\rm MeV}$ antineutrino as functions of the radius $r$ but now with the
value of $\theta_{13}$ set to $\sin^{2}(2\theta_{13})=4\times 10^{-6}$.
The instantiation of the density is shown in figure (\ref{fig:1}).
Again, the turbulence region is identified by the vertical dashed lines and the
evolution through the base profile as the dashed line. 
The hierarchy  is inverted and the fluctuation amplitude is $C_{\star}=0.1$.
\label{fig:7}}
\vspace*{2mm}
\end{figure}
\begin{figure}
\vspace*{2mm}
\includegraphics[clip, width=\linewidth]{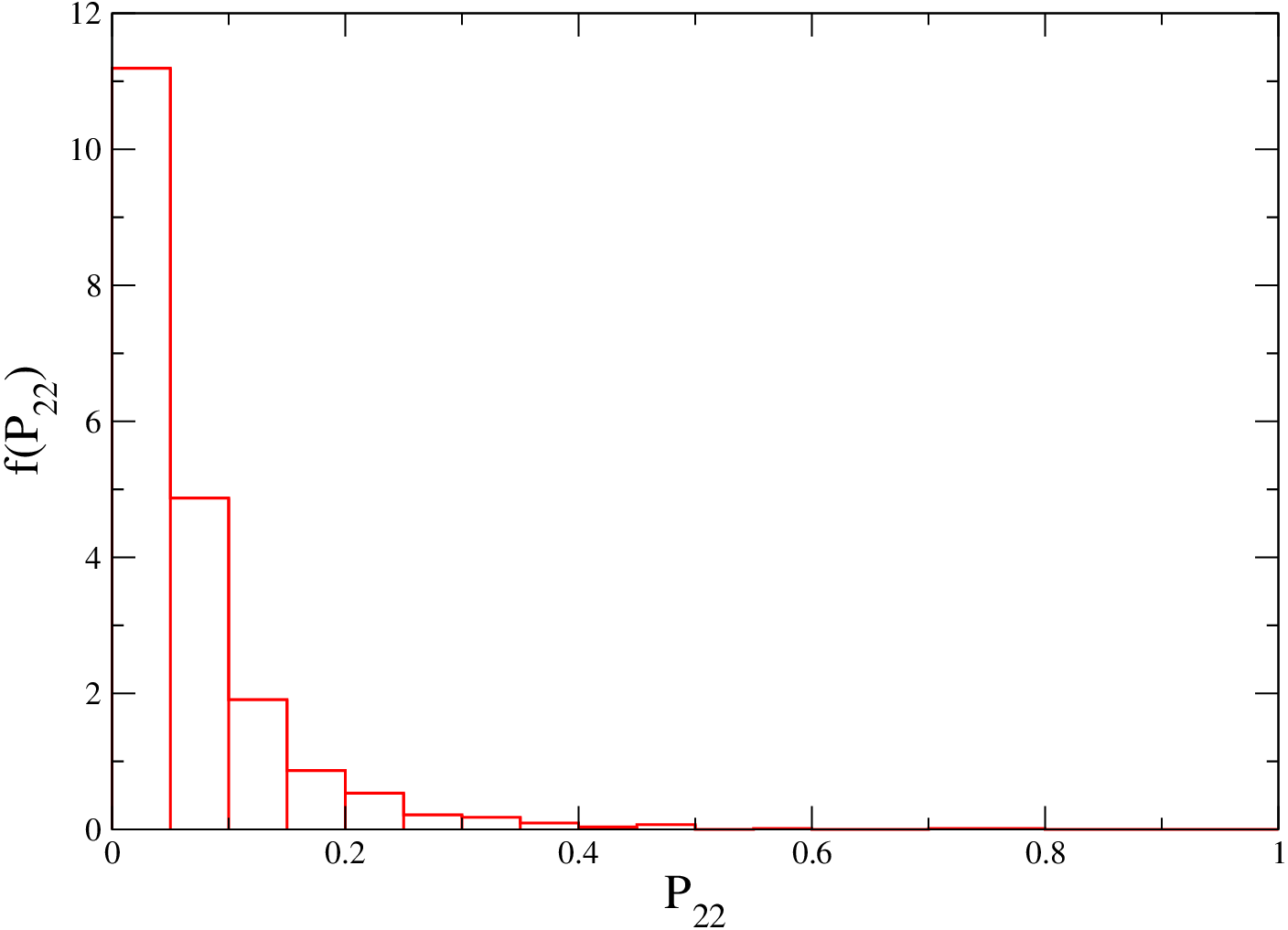}
\caption{A normalized frequency histogram of 1687 calculations of
$P_{22}$ for a $E=25\;{\rm MeV}$
neutrino with a fixed underlying profile and $C_{\star}=0.1$. Here
$\theta_{13}$ is set to $\sin^{2}(2\theta_{13})=4\times 10^{-6}$ and the
hierarchy normal.\label{fig:8}}
\vspace*{5mm}
\end{figure}
Investigating how turbulence effects are modified when one decreases
$\theta_{13}$ is important because the third neutrino mixing angle may not turn
out to be close to the present Chooz limit and able to be measured in the
forthcoming reactor and first superbeam experiments. In a core-collapse
supernova without turbulence, the non-adiabaticity of the H resonances of 
profiles from spherically symmetric simulations becomes large for values of
$\theta_{13}$ smaller than the Dighe-Smirnov threshold of $\sin^2 2\theta_{13} \sim 10^{-5}$
\cite{Dighe:1999bi}. But when we include turbulence the limiting behavior that survival probabilities
for the states that mix at the H resonance 
should tend to zero as $\theta_{13} \rightarrow 0$ need not occur because
the smallness of any non-zero $\theta_{13}$ can always be compensated by
increasing the amplitude and/or extent of 
the matter fluctuations. However the fluctuation amplitude cannot be increased
beyond reason and the region where turbulence occurs is limited to the space 
between the shocks in our model. It is due to these `upper limits' on the compensating variables that indeed we can 
expect to reach the limiting behavior of non-adiabatic evolution for the H resonant mixing states 
as $\theta_{13} \rightarrow 0$ at some point but note it will not necessarily be at the same
value as the Dighe-Smirnov threshold.
This expectation is borne out when one investigates the effect of turbulence for small values of
$\theta_{13}$ since one observes there can be significant, i.e. greater than $\mathcal{O}(1\%)$, 
changes in the neutrino survival probability for values of
$\sin^{2}2\theta_{13}=4\times 10^{-6}$, as shown in figure (\ref{fig:7}). For this
particular figure the hierarchy is inverted so the H resonance 
mixes the antineutrino states $\bar{\nu}_{1}$ and $\bar{\nu}_{3}$, for a normal
hierarchy mixing is between ${\nu}_{2}$ and ${\nu}_{3}$ but the essential
behavior is the same for both.

As for the case of large $\theta_{13}$, each instantiation of $F$ leads to a
different final state and again we can 
construct the distribution of the possible final states by Monte Carlo  methods.
The distribution of the final state $P_{22}$ for the case of 
a normal hierarchy, $E=25\;{\rm MeV}$, $\sin^{2}(2\theta_{13})=4\times 10^{-6}$
and $C_{\star}=0.1$ is shown in figure (\ref{fig:8}).
We see that the distribution is not a delta-function at zero corresponding to
non-adiabatic propagation but rather appears to have an exponential behavior.

\begin{figure}[t]
\vspace*{5mm}
\includegraphics[width=\linewidth]{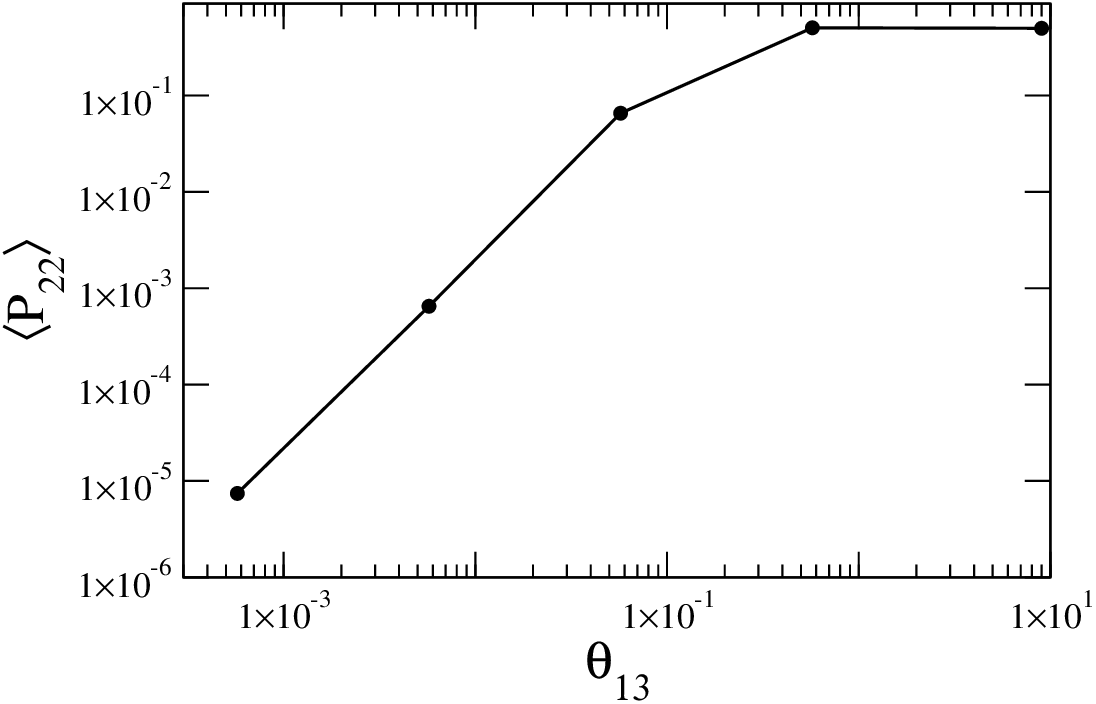}
\caption{Mean value of the second matter eigenstate for $E=25\;{\rm MeV}$
neutrinos as a function of third neutrino mixing angle, in degrees, for the case of normal
hierarchy and $C_{\star}=0.1$.  \label{fig:9}}
\vspace*{2mm}
\end{figure}
With figure (\ref{fig:9}) we summarize the entire behavior of the turbulence
effect as a function of $\theta_{13}$ for a fixed matter density fluctuation
amplitude $C_{\star}=0.1$. For `large' angles - above the Dighe-Smirnov
threshold- we are in the depolarized limit for this value of $C_{\star}=0.1$,
below the threshold we are in a scaling regime where $P_{22}\propto
\theta_{13}^{2}$. Setting a limit that observable changes require $\langle P_{22}\rangle \geq 0.01$ 
for present neutrino detectors then we see that turbulence induces observable sensitivity to $\theta_{13}$ for values above $0.02^{\circ}$ or, 
equivalently, $\sin^{2}(2\theta_{13}) \gtrsim 4\times 10^{-7}$.  

\subsection{Three flavor resonant and non-resonant mixing}

To complete our study of turbulence effects we look at large amplitude
fluctuations, their breaking of HL factorization and the impact in the channels
where no MSW resonance(s) occur. We take the case of normal hierarchy as an
example so the non-L-resonant and non-H-resonant transitions occur for
antineutrinos.
\begin{figure}[b]
\vspace*{2mm}
\includegraphics[width=\linewidth]{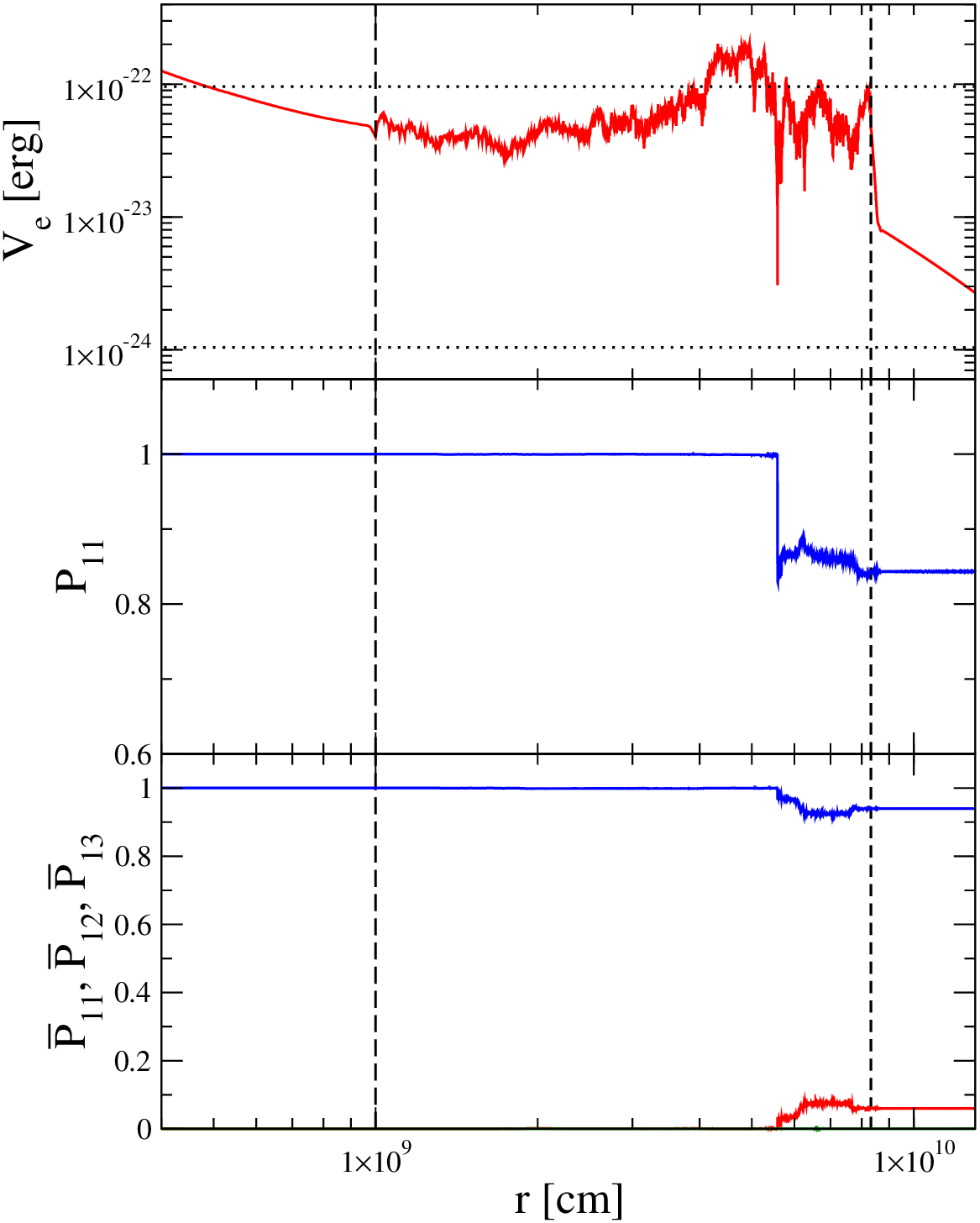}
\caption{Impact of turbulence on the first matter eigenstates for the neutrino
survival probability (middle) and for the anti-neutrino survival and appearance
probabilities (bottom). The results correspond to normal hierarchy,
$\sin^{2}(2\theta_{13})=4\times 10^{-4}$, 
$C_{\star}^2=0.34$ and $E = 25\;{\rm MeV}$. The top panel shows the matter
density profile with noise and again the
vertical dashed lines indicate the turbulence region while the upper horizontal
line is the H resonance and the lower the L resonance for this 
particular neutrino energy computed using the two-flavor formula. 
\label{fig:10}}
\vspace*{2mm}
\end{figure}
In figure (\ref{fig:10}) we present the turbulence density profile and the
corresponding neutrino and anti-neutrino probabilities for the case of
$\sin^{2}(2\theta_{13})=4\times 10^{-4}$ and $C_{\star}^2=0.34$. The amplitude
of the fluctuations here is so large that HL factorization is broken. This can
be seen from the evolution of $P_{11}$ in the middle panel and noting that it
occurs at the point where $V_e$ in the top panel approaches the L resonance at
$r \sim 60,000\;{\rm km}$. 
Also note that that breaking HL factorization does not depend upon the
hierarchy. In the lower panel we see that simultaneously there is a sudden change in the
anti-neutrino transition probabilities $\bar{P}_{11}$ and $\bar{P}_{12}$ at the point where the 
L resonance occurs but we
do not observe any variation of $\bar{P}_{13}$ which remains at zero for all
$r$ in this case.
\begin{figure}
\vspace*{2mm}
\includegraphics[width=\linewidth]{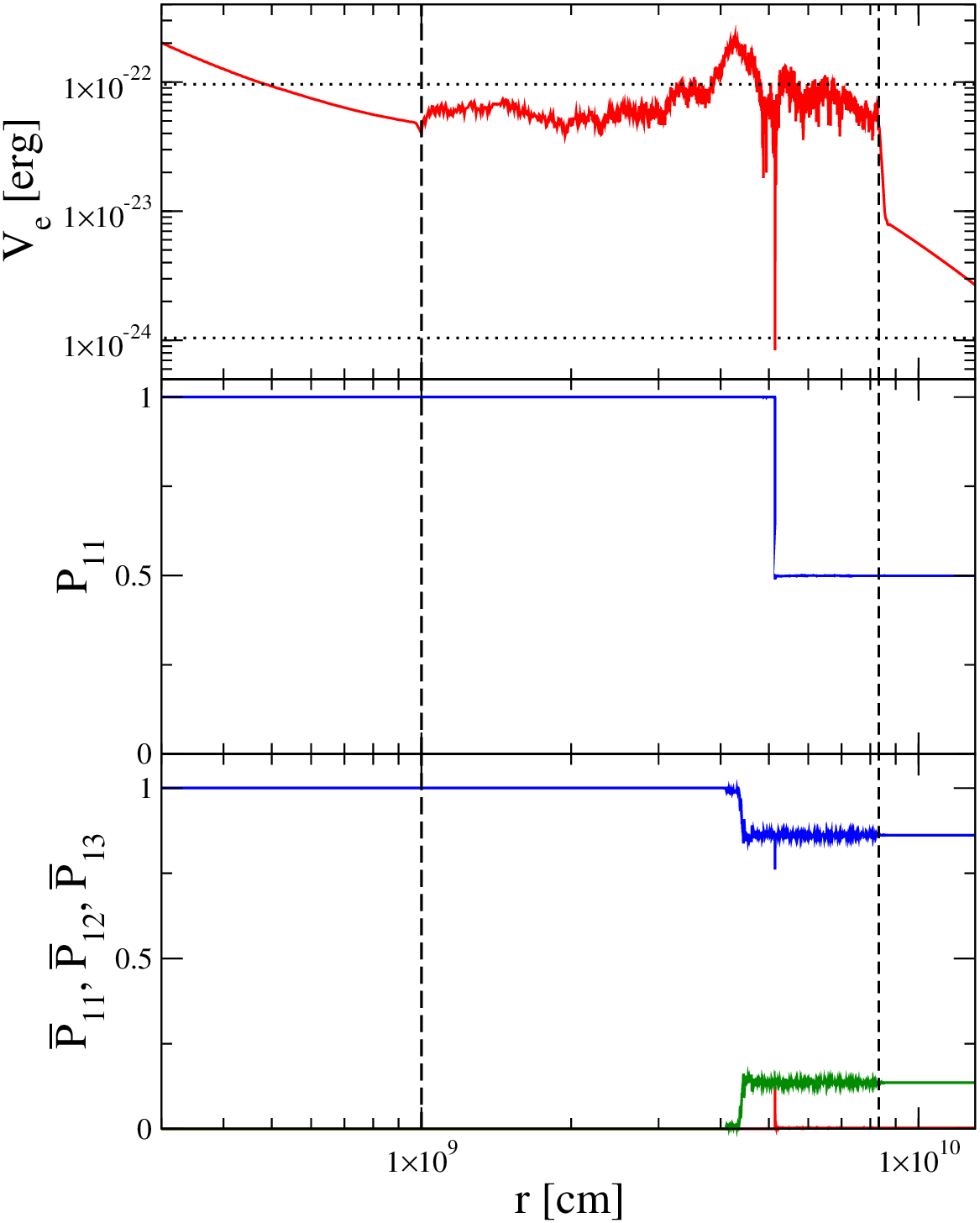}
\caption{Impact of turbulence on the first matter eigenstates for the neutrino
survival probability (middle) and for the anti-neutrino survival and appearance
probabilities (bottom).The results correspond to normal hierarchy,
$\theta_{13}=9^{\circ}$, 
$C_{\star}^2=0.1$ and $E = 25\;{\rm MeV}$. The top panel shows the matter
density profile with noise and, like figure (\ref{fig:10}), the
vertical dashed lines indicate the turbulence region while the upper horizontal
line is the H resonance and the lower the L resonance for this 
particular neutrino energy computed using the two-flavor formula. \label{fig:11}}
\vspace*{2mm}
\end{figure}
In order to observe non-trivial evolution of $\bar{P}_{13}$ we need to increase $\theta_{13}$ close to the Chooz limit. An example 
with much larger $\theta_{13}$ is shown in 
figure (\ref{fig:11}) for the case of $\theta_{13}=9^{\circ}$ and
$C_{\star}^2=0.1$. Again HL is broken, this time at $r \sim 50,000\;{\rm km}$,
causing a sudden decrease in $P_{11}$ and simultaneous jumps in $\bar{P}_{11}$
and $\bar{P}_{12}$ but we also notice from the lower panel that $\bar{P}_{13}$ is already non-zero
by this point; mixing between states 
$\bar{\nu}_1$ and $\bar{\nu}_3$ had commenced when $V_e$ approached the H
resonance at $r \sim 40,000\;{\rm km}$ as shown in the top panel.
Both these non resonant mixing effects can be understood through the
antineutrino non-adiabaticity parameters $\bar{\Gamma}_{12}$ and
$\bar{\Gamma}_{13}$ which depend upon the derivatives of the matter mixing
angles $\tilde{\theta}_{12}$ and $\tilde{\theta}_{13}$ respectively. For both,
the matter mixing angle varies between the high density limit of zero to its
value in vacuum at resonances. For $\tilde{\theta}_{12}$ this occurs at L
resonances, for $\tilde{\theta}_{13}$ at H resonances. Thus we expect that the
larger the value of the vacuum angle the greater the derivative and the
non-adiabaticity parameters $\bar{\Gamma}_{12}$ and $\bar{\Gamma}_{13}$ and,
therefore, the greater the mixing between the matter eigenstates $\bar{\nu}_1$
and $\bar{\nu}_2$ or $\bar{\nu}_1$ and $\bar{\nu}_3$. The difference between the mixing of neutrinos 
states $\bar{\nu}_1$-$\bar{\nu}_2$ and $\bar{\nu}_1$-$\bar{\nu}_3$ is that the vacuum value of ${\theta}_{12}$ is well known and large while
the value of ${\theta}_{13}$ is neither large nor known. 
This effect of of non-resonant mixing between states $\bar{\nu}_1$ and
$\bar{\nu}_3$ is the same as found in Ref.\cite{Fogli:2006xy} where it is  also
shown that the turbulence impact on the non-resonant neutrino channel disappears
for a decreasing value of $\theta_{13}$ though we would like to emphasize that
our findings are for a very different set-up than the one in \cite{Fogli:2006xy}. It is important
to note that the non-resonant effect from $\tilde{\theta}_{12}$ does not
depend upon the hierarchy, it will always occur for the anti-neutrinos and never
for the neutrinos because the sign of $\delta m_{12}^{2}$ is known. In contrast
the non-resonant effect from $\tilde{\theta}_{13}$ will switch to the neutrinos
if we consider an inverted hierarchy.

\begin{figure}[t]
\vspace*{2mm}
\includegraphics[clip=true, width=\linewidth]{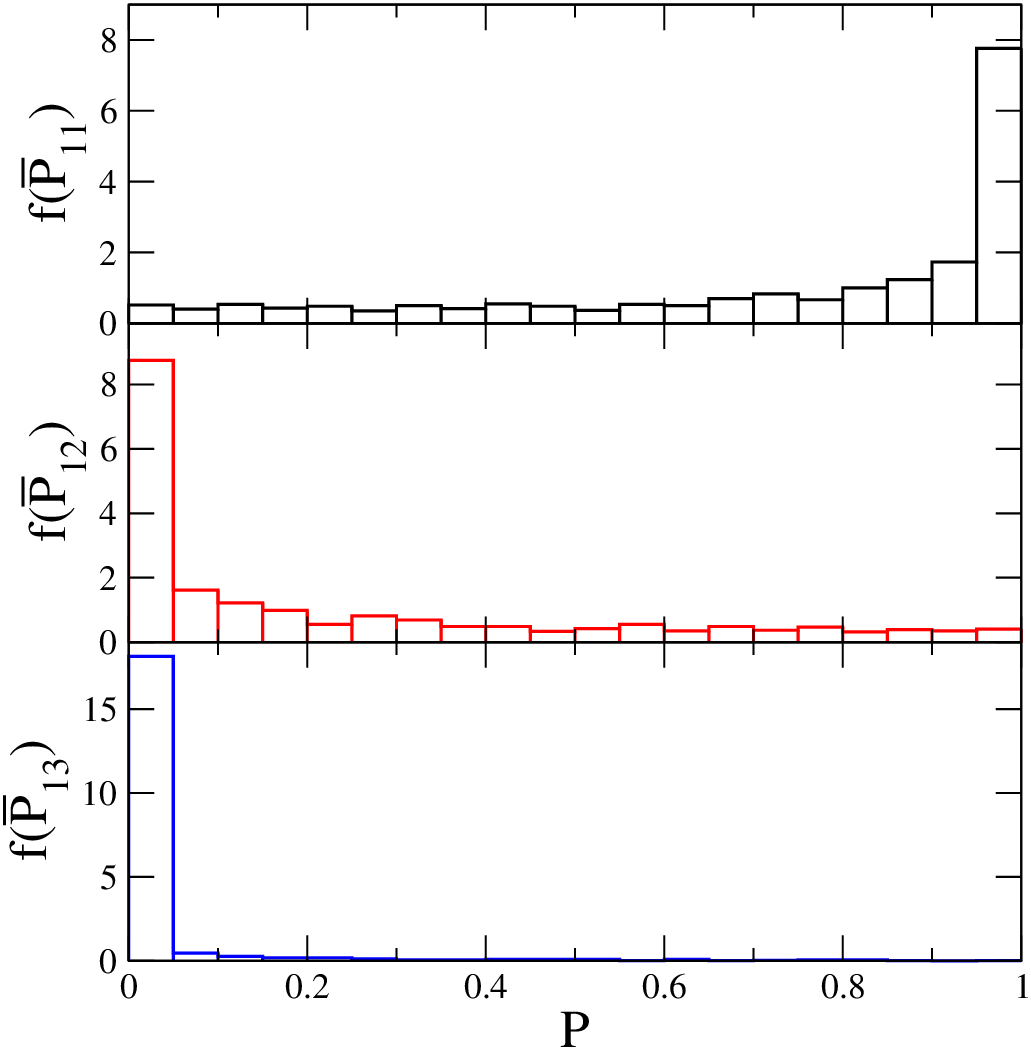}
\caption{Frequency distribution for the anti-neutrino probabilities for a normal
hierarchy, $\theta_{13}=9^{\circ}$, $C_{\star}^2=0.2$ and $E = 25\;{\rm
MeV}$.\label{fig:12}}
\vspace*{2mm}
\end{figure}
Thus we find two effects from turbulence in the non resonant channels. One from
broken HL factorization causing non-L-resonant transitions and the other from the
constant transitions from the high density to vacuum values of $\theta_{13}$. As
with the case of small $\theta_{13}$ and the H resonant channel, the
distributions of the non resonant transition probabilities for an ensemble have
an exponential like behavior. An example of the distributions for the particular
case of for a normal hierarchy, $\theta_{13}=9^{\circ}$, $C_{\star}^2=0.2$ and
$E = 25\;{\rm MeV}$ is shown in figure (\ref{fig:12}).
\begin{figure}
\vspace*{2mm}
\includegraphics[clip=true, width=\linewidth]{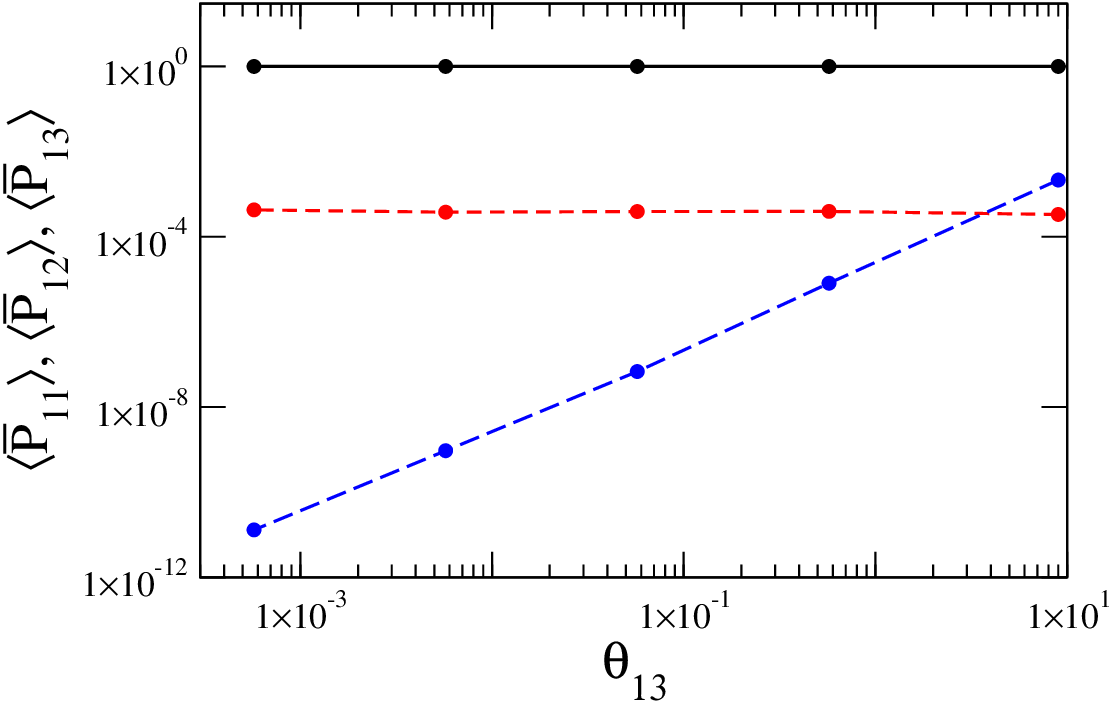}
\caption{Variation of the mean anti-neutrino survival probabilities $\langle
\bar{P}_{11} \rangle$ (solid), $\langle \bar{P}_{12} \rangle$ (short dashed) and
$\langle \bar{P}_{13} \rangle$ (long dashed) as a function of the vacuum mixing
angle $\theta_{13}$ for the case of a normal hierarchy, $C_{\star}=0.1$ and $E =
25\;{\rm MeV}$. The symbols indicate the values of $\theta_{13}$ where averages
were calculated.\label{fig:13}}
\vspace*{2mm}
\end{figure}
\begin{figure}
\vspace*{2mm}
\includegraphics[clip=true, width=\linewidth]{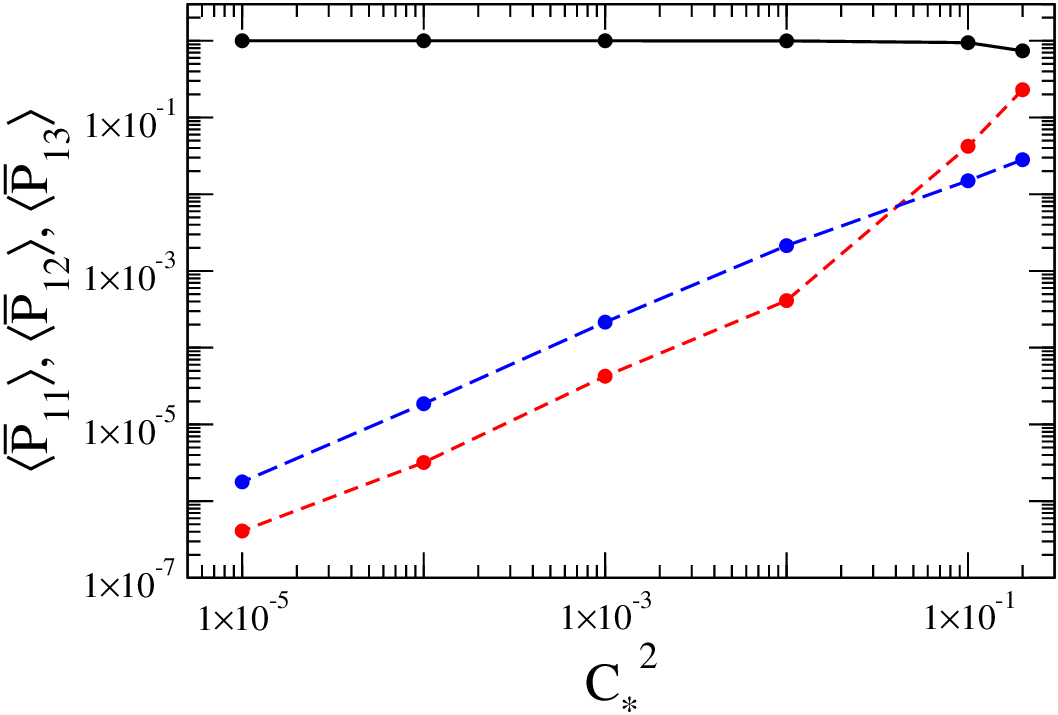}
\caption{Variation of the mean anti-neutrino survival probabilities $\langle
\bar{P}_{11} \rangle$ (solid), $\langle \bar{P}_{12} \rangle$ (short dashed) and
$\langle \bar{P}_{13} \rangle$ (long dashed) as a function of the fluctuation
amplitude $C_{\star}$ for a vacuum mixing angle $\theta_{13}=9^{\circ}$, the
case of a normal hierarchy, and $E = 25\;{\rm MeV}$. The symbols indicate the
values of $C_{\star}$ where averages were calculated.\label{fig:14}}
\vspace*{2mm}
\end{figure}
As explained above, the average values of the distributions scale differently with ${\theta}_{13}$
and the variation of the turbulence effect with the third neutrino mixing angle
is shown in figure (\ref{fig:13}). While the first matter eigenstate survival
probability and appearance probability to the second matter eigenstate $\langle
\bar{P}_{12} \rangle$ are constant the mixing with the third matter
eigenstate, $\langle \bar{P}_{13} \rangle$, scales as $\theta^{2}_{13}$ and so
disappears as $\theta_{13}$ decreases. We also observe that the mixing between
states $\bar{\nu}_1$ and $\bar{\nu}_3$ surpasses that between $\bar{\nu}_1$ and
$\bar{\nu}_2$ only for ${\theta}_{13} \geq 3^{\circ}$. Finally, the behavior of
$\langle \bar{P}_{11} \rangle$, $\langle \bar{P}_{12} \rangle$ and $\langle
\bar{P}_{13} \rangle$ as a function of the fluctuation amplitude $C_{\star}^{2}$
is shown in Fig.(\ref{fig:14}). At small amplitudes both $\langle \bar{P}_{12}
\rangle$ and $\langle \bar{P}_{13} \rangle$ scale as $C_{\star}^{2}$ but as
$C_{\star}^{2}$ increases $\langle \bar{P}_{12} \rangle$ appears to increase its
sensitivity to this parameter.

\section{Conclusions}
We have performed a detailed investigation of turbulence effects after solving
the evolution equation for neutrinos passing through turbulent density profiles
created by adding matter density fluctuations to a 1D density
profile obtained from hydrodynamical simulations. Our treatment differs from
most previous studies since we do not solve the neutrino evolution equations for
the averaged probabilities but rather for the neutrino amplitudes. The results
have highlighted new turbulence effects for that period when the shocks are in
the region of the H resonance for neutrino
energies of order $\mathcal{O}(10\;{\rm MeV})$ and can be summarized as
follows.
\begin{itemize}
\item
For large third neutrino mixing angle but fluctuation amplitudes below
$C_{\star}^{2} \sim 0.1$ we find turbulence effects only appear in the channel
with the H resonance i.e. the neutrinos for a normal hierarchy, the
antineutrinos for an inverted hierarchy. More specifically we find that if the
matter density fluctuation amplitudes are tiny, no turbulence effects appear as
expected; for intermediate values, $10^{-10} \lesssim C_{\star}^{2} \lesssim
10^{-2}$ the neutrino matter survival probabilities are dominated by the phase
effects between the pre-existing resonances of the profile but with fluctuating
relative phase that are sufficient to produce a wide variation of the final
states; for larger values of the matter density fluctuation amplitude,  $10^{-2}
\lesssim C_{\star}^{2} \lesssim 10^{-1}$, turbulence effects begin to dominate
and lead to two-flavor depolarization. The sensitivity to amplitudes as small $0.001\%$ would 
apparently indicate that turbulence effects in this channel are inevitable but, at the 
same time, observationally turbulence effects will be difficult to distinguish from phase effects once one takes 
into account time and energy binning of a signal, as pointed out in Galais \emph{et al.} \cite{Galais:2009wi}.

\item 
When the matter density fluctuation amplitude is large, $C_{\star}^{2} \gtrsim
0.1$, we have found a sensitivity in the H resonant channel to the third
neutrino mixing angle below the Dighe-Smirnov threshold down to $\sin^{2}(2\theta_{13}) \gtrsim 4\times 10^{-7}$.
\item
Large amplitude matter density fluctuations, $C_{\star}^{2} \gtrsim 0.1$, lead
to broken HL independent of $\theta_{13}$ and the hierarchy. This leads to
resonant mixing between states $\nu_1$ and $\nu_2$ \emph{and} non-resonant
mixing between states $\bar{\nu}_1$ and $\bar{\nu}_2$.
\item
Once again, for large amplitude matter density fluctuations, $C_{\star}^{2}
\gtrsim 0.1$, we find that non-resonant mixing may occur due to the fluctuations
of the matter mixing angle $\tilde{\theta}_{13}$. The degree of mixing is a
function of $\theta_{13}$ with a mean value that scales as $\theta_{13}^{2}$.
The channel where this effect appears depends upon the hierarchy appearing in
the antineutrinos for a normal hierarchy where it leads to mixing between
$\bar{\nu}_1$ and $\bar{\nu}_3$ and in the neutrinos for an inverted hierarchy
where it causes mixing between states $\nu_2$ and $\nu_3$.
\end{itemize}
When the matter density fluctuations are small turbulence can only cause two
flavor mixing. When the amplitude is large three flavor mixing may occur in the
neutrinos and antineutrinos simultaneously but, depending upon the hierarchy and
$\theta_{13}$, not necessarily to the same degree. Let us summarize our findings
in this case of large amplitude fluctuations:
\begin{itemize}
\item
In the case of normal hierarchy: for $\theta_{13}$ above the Dighe-Smirnov
threshold strong mixing occurs in the neutrinos and 3-flavor depolarization may
be approached. On the other hand, if $\theta_{13}$ is close to the Chooz limit,
3-flavor mixing of the antineutrinos occurs, due to the two non-resonant
effects; otherwise only two-flavor mixing takes place because of the
non-resonant effect from $\tilde{\theta}_{12}$ fluctuations. 
When $\theta_{13}$ is below the Dighe-Smirnov threshold, the neutrinos still
exhibit three flavor mixing due to broken HL factorization and the residual
sensitivity to $\theta_{13}$, that scales as $\theta_{13}^{2}$.
\item In the case of inverted hierarchy: for large fluctuations and
$\theta_{13}$ close to the Chooz limit, 3-flavor mixing occurs in the neutrinos,
due to broken HL factorization, and the non-resonant effect from
$\tilde{\theta}_{13}$ fluctuations which will turn into 2-flavor non-resonant
mixing of $\nu_1$ and $\nu_2$ as $\theta_{13}$ decreases. Three flavor mixing of
the antineutrinos takes place for $\theta_{13}$, above the Dighe-Smirnov
threshold, but it is unlikely that depolarization is reached because the mixing
between $\bar{\nu}_1$ and $\bar{\nu}_2$ is non-resonant. 
For $\theta_{13}$ below the Dighe-Smirnov threshold, some residual three-flavor
mixing will remain, but it will be small, because now the two turbulence effects
are perturbative in this channel.
\end{itemize}

Several aspects related to turbulence remain open. While our treatment of
turbulence can already be considered, in many 
respects, realistic it is clear that future studies are awaited where matter
density fluctuations are drawn from successful core-collapse supernova
simulations at many different epochs. Also, until we can calibrate these
one-dimensional studies the actual turbulence effects we should expect is
unknown. 

Lastly, one important question that needs further investigation is the extent of
the `smearing' or `blurring' of turbulence features in a burst signal. Smearing can occur due 
several reasons. First there is the smearing due to the finite size of the source.
The neutrinos we shall receive from the next Galactic supernova will not travel along a single 
line of sight so the neutrino flux at a given arrival time and neutrino energy 
(ignoring time of flight issues \cite{1998PhRvD..58i3012B}) is `averaged' over a
column of $\sim 20\;{\rm km}$ i.e.\ the apparent diameter of the proto-neutron star source. 
Neutrinos emitted from different locations upon the source will pass through different small scale, $\lesssim 20\;{\rm km}$,
fluctuations but the turbulence on large scales will be the same for all the neutrinos we receive in the burst. 
Second, a smearing in energy will occur in the detector due to the intrinsic energy resolution of the
detection reaction and the extrinsic resolution of the detector. If the combined effect of intrinsic and extrinsic 
resolutions leads to a smearing over an energy range $\sigma$ then features on scales smaller than $\sigma$ will 
be unobservable. Finally, smearing in both time and energy will also be introduced because of the necessity of binning the detected events.
As with the detector response, short time scale changes in the flux or features with small energy widths will be smeared reducing 
their observability. Studying how all these effects impact the observability of the turbulence effects pointed out
in the present paper is not trivial and will be the object of future investigations. 
\cite{JKCVnext}.

\textit{}

\end{document}